\documentclass[twocolumn,showpacs,showkeys,preprintnumbers,amsmath,amssymb,nofootinbib]{revtex4}
\usepackage{graphicx}% Include figure files
\usepackage{dcolumn}% Align table columns on decimal point
\usepackage{bm}% bold math
\begin{document}
%\usepackage{graphics}
%\begin{center}{
\title{\bf Generalized  Emission Functions for Photon Emission from Quark-Gluon Plasma }
%\\
\author{S.  V.  Suryanarayana\footnote{  In  nuclear  physics journals and arxiv
listings, my name used to appear as S.V.S.  Sastry.
Hereafter (by deleting Sastry),  my name will appear as, S.V. Suryanarayana.}}
\email{snarayan@barc.gov.in}
\email{suryanarayan7@yahoo.com}
\affiliation{  Nuclear Physics Division, Bhabha Atomic Research Centre, Trombay, Mumbai
400 085, India}
\begin{abstract}{
The  Landau-Pomeranchuk-Migdal  effects on photon emission from the quark
gluon plasma have been studied as a function of photon mass, at a  fixed
temperature of the plasma. The integral equations for the transverse vector
function   (${\bf   \tilde{f}(\tilde{p}_\perp  )}$)  and  the  longitudinal
function  ($\tilde{g}({\bf \tilde{p}_\perp } )$)  consisting  of  multiple   scattering
effects are solved by the self consistent iterations method and also by the
variational  method  for  the variable set \{$p_0,q_0,Q^2$\}, considering the
bremsstrahlung and the $\bf aws$ processes. We define four new dynamical scaling
variables,  $x^b_T$,   $x^a_T$,   $x^b_L$,   $x^a_L$, 
for bremsstrahlung and {\bf aws} processes and analyse  the  transverse   and   longitudinal   components   as   a  function  of
\{$p_0,q_0,Q^2$\}. We  generalize  the  concept  of  photon  emission
function  and we define four new emission functions for massive photon emission  represented by
$g^b_T$,   $g^a_T$,   $g^b_L$,   $g^a_L$. 
These   have    been constructed  using  the  exact numerical solutions of the integral equations.
These  four  emission functions   have   been parameterized  by  suitable simple empirical  fits.
 In terms of these empirical emission functions, the virtual photon emission from  quark gluon plasma  reduces
to  one  dimensional  integrals  that  involve  folding  over the empirical
$g^{b,a}_{T,L}$ functions with appropriate quark distribution functions and
the  kinematic  factors.  Using  this  empirical  emission  functions,   we
calculated  the  imaginary  part  of  the  photon  polarization tensor as a function of photon mass and energy.
}\end{abstract}
%%%%%%--------------------------------------------------------------------
\maketitle
\par
\noindent
Quark gluon plasma (QGP) state is expected to be formed in the relativistic
heavy ion collisions. In order to identify the plasma or a de-confined state,
one  needs  to study the physical processes in quark matter,  that  can  distinctly  and conclusively identify
this  state  of  matter, such as parton energy loss leading to jet suppression mechanism.  In this context, electromagnetic processes such as
photons and dilepton emission are also considered as  important  signals.
Photons  and  dileptons  are  emitted  at  various  stages  during  plasma
evolution, for an overview one may see  \cite{peitz,gale,rapp}  and
the  references  therein. In depth study of photon emission processes in quark-gluon plasma were reported 
\cite{kapusta,bair} including processes also from hot hadron gas  \cite{kapusta}.  
In the formalism of  hard  thermal  loops \cite{braaten} (HTL) effective theory, 
the  processes  of  bremsstrahlung \cite{brem} and a crossed
process of off-shell annihilation called ${\bf  aws}$  \cite{bremaws,ktkl} contribute to photon emission at the two loop level. 
These two processes contribute at the leading order $O(\alpha\alpha_s)$ owing to the  collinear
singularity   that   is   regularized   by  the  effective  thermal  masses. 
Higher loop multiple scatterings having a ladder topology also contribute at the
same    order     as     the     one     and     two     loop     processes
\cite{arnold1,arnold2}.   These   higher  loop
rescatterings, each giving finite decoherent correction to the two loop processes,  need 
to    be    resummed.   This   resummation   effectively   implements   the
Landau-Pomarenchk-Migdal (LPM) effects \cite{landau1,landau2,migdal} 
arising  due  to  rescattering  of
quarks in the medium during the photon formation time. 
This results in an integral equation of a transverse
vector function for the real photons.  The photon emission rates are
suppressed   owing   to   the  LPM  effects  \cite{arnold1,arnold2}.
The  LPM modification  of  the  photon  spectrum  is  important  at  very low photon
energies or at  high  photon  energies.  For  example,  the  bremsstrahlung
radiation  is  strongly  suppressed  by  almost  $~80\%$  at very low $q_0/T$
values, whereas the photon emission  from  $\bf  aws$  falls  strongly  for
higher  $q_0/T$  values \cite{arnold2}. Thus the rescattering corrections  modify the two
loop contributions for bremsstrahlung and $\bf aws$ processes  at  opposite
ends  of  the photon spectrum for real photons. 
\section{Emission of  Real Photons}
The photon production rates
from bremsstrahlung  and  the  $\bf  aws$  processes 
considering the LPM effects  are  estimated by using Eq.\ref{photrate} \cite{arnold2}.
\begin{eqnarray}
{\cal R}_{b,a}&=& \frac{80\pi T^3\alpha\alpha_s}{(2\pi)^3 9\kappa}\int dp_0
\left[\frac{p_0^2+(p_0+q_0)^2)}{p_0^2(p_0+q_0)^2)}\right]  \left[n_f(q_0+p_0) \right. \nonumber \\
&&\left. (1-n_f(p_0))\right]~\int \frac{\bf d^2\tilde{p}_\perp}{(2\pi)^2}  2{\bf \tilde{p}_\perp \cdot\Re\tilde{f}(\tilde{p}_\perp)}
\label{photrate}
\end{eqnarray}
\begin{eqnarray}
2{\bf \tilde{p}_\perp}&=& i\delta \tilde{E}({\bf \tilde{p}_\perp}) {\bf \tilde{f}}({\bf \tilde{p}_\perp}) +  \nonumber \\
&& \int\frac{d^2{\bf \tilde{\ell}_\perp}}{(2\pi)^2}\tilde{C}({\bf \tilde{\ell}_\perp})
\left[{\bf \tilde{f}}({\bf \tilde{p}_\perp})
-{\bf \tilde{f}}({\bf \tilde{p}_\perp+\tilde{\ell}_\perp})\right]
\label{amy}
\end{eqnarray}
\begin{eqnarray}
\delta \tilde{E}({\bf \tilde{p}_\perp})&=&
\frac{q_0T}{2p_0(q_0+p_0)}\left[\tilde{p}_\perp^2+\kappa \right]
\label{amydelta}
\end{eqnarray}
\begin{eqnarray}
{\cal R}_{b,a}&=&{\cal C}_k \int dp_0
\left[p_0^2+(p_0+q_0)^2)\right]\left[n_f(q_0+p_0 ) \right. \nonumber \\
&& \left. (1-n_f(p_0 ))\right]C_g g(x)
\label{gxprc}
\end{eqnarray}
\begin{eqnarray}
g(x)&=&g(p_0,q_0,\kappa,T) \\
x&=&\frac{1}{\kappa_0}\frac{q_0T}{p_0(p_0+q_0)} \\
{\cal C}_k&=&\frac{40\alpha\alpha_sT}{9\pi^4q_0^2} \\ 
\kappa_0=\frac{M_\infty^2}{m_D^2} ~\mbox{and}~C_g&=&\frac{\kappa}{T}\left(\frac{\kappa_0}{\kappa}\right)^{-0.40}
\end{eqnarray}
\noindent
$\Re{\bf  \tilde{f}(\tilde{p}_\perp  )}$  in Eq.\ref{photrate} is the real
part of a transverse vector function which consists of the LPM effects  due
to  multiple  scatterings.  This can be taken as transverse momentum vector
${\bf (\tilde{p}_\perp )}$ times a scalar function of  transverse  momentum
${\tilde{p}_\perp  }$.  The  sign  ~$\tilde{}$~  denotes  the dimensionless
quantities  in  units of Debye mass $m_D$ as defined in \cite{arnold2}. The
function  ${\bf  \tilde{p}_\perp\cdot\Re\tilde{f}(\tilde{p}_\perp  )}$   is
determined by the collision kernel ($\tilde{C}({\bf \tilde{q}_\perp})$) in
terms  of  the  AMY  integral  equation (Eq.\ref{amy})  and the energy function
$\delta E$  (Eq.\ref{amydelta}). This was solved in \cite{arnold2} using the variational approach.
We  simplified  this  variational  approach  and  extended to finite baryon
density case \cite{svs1}. Moreover, we reported that the complex LPM  effects
can  be  very  well  reproduced by introducing the photon emission function
$g(x)$ of dynamical variable $x$ \cite{svsprc}.  In  terms  of  the  single
variable function  $g(x)$ , the photon emission rates are estimated as shown
in Eq.\ref{gxprc}.  A figure for $g(x)$ function for real photons will be presented  later
(Fig.\ref{gxbtgx}).
\section{Emission of Virtial Photons}
Processes that contribute to virtual photon emission in QGP  
at  $\alpha\alpha_s$  order \cite{alther} and the  higher order corrections  \cite{thoma} were reported.
The processes $q\bar{q}\rightarrow g\gamma^*$ , and  $qg\rightarrow q\gamma^*$ contribute to photon  
polarization tensor at the order of  $\alpha\alpha_s$ and appear as
the  one  loop processes in HTL method. 
Further considering LPM effects, photon  emission  from  QGP  
as  a  function of photon mass was also reported \cite{lpmdilep}.
For  the  case  of  virtual  photons, these multiple scatterings modify the
imaginary part of self energy as a function of photon energy  and  momentum
both.  It  has  been shown that the rescatterings modify the self energy  at low $Q^2$, more
importantly around the tree level threshold,  $Q^2 = 4M_\infty^2$ \cite{lpmdilep}.  
The dilepton emission rates are estimated in terms of  the imaginary part of 
retarded photon polarization tensor,   given by \cite{lpmdilep}.
\begin{equation}
\frac{dN_{\ell^+\ell^-}}{d^4xd^4Q} = \frac{\alpha_{EM}}{12\pi^4Q^2(e^{q0/T}-1)} \Im\Pi_{R\mu}^\mu(Q)
\end{equation}
\par
\noindent
For  the  case  of  virtual  photon  emission  having small virtulaity, the
transverse vector function is  determined  by  the  Eq.\ref{agmz-t}     and     the     energy    transfer    $\delta    {E}(({\bf
{p}_\perp},p_0,q_0,Q^2)$ given in  Eq.\ref{agmz-de}  \cite{lpmdilep}.  For
virtual  photons,  the  coupling  of  quarks  to  longitudinal mode must be
considered. This results in a scalar function of ${\bf p_\perp}$  which  is
determined  by  an  integral  equation  and  collision  kernel in  Eq.\ref{agmz-l}. For simplicity, henceforth I will refer to 
Eq.\ref{agmz-l} as AGMZ equation. An analytical form for the collision kernel was
given in \cite{kernel}, owing to which the photon emission rate calculations became easy.
The $\delta{E}$ is an energy denominator  and  can
be  interpreted as inverse formation time of the photon. For
the case of massive photon emission, this energy denominator is modified by
replacing $M_\infty^2 \rightarrow M_{\mbox{eff}}=M_\infty^2+\frac{Q^2}{q_0^2}{2p_0r_0}$.
For  $Q^2>4M_\infty^2$,  this  M$_{\mbox{eff}}$  can  vanish or even become
negative.
\begin{eqnarray}
2{\bf {p}_\perp}&=&i\delta {E}({\bf {p}_\perp},p_0,q_0,Q^2) {\bf {f}}({\bf {p}_\perp}) +  g^2C_FT \nonumber \\
&& \int \frac{d^2{\bf{\ell}_\perp}}{(2\pi)^2}{C}({\bf {\ell}_\perp}) \left[{\bf {f}}({\bf {p}_\perp})
-{\bf {f}}({\bf {p}_\perp+{\bf\ell}_\perp})\right]
\label{agmz-t}
\end{eqnarray}
\begin{equation}
\delta E({\bf{p}_\perp},p_0,q_0,Q^2)=
\frac{q_0}{2p_0r_0}\left[{\bf{p}_\perp^2}+M_{\mbox{eff}}^2 \right]
\label{agmz-de}
\end{equation}
\begin{eqnarray}
2{\bf \sqrt{|p_0r_0|}}&=&i\delta {E}({\bf{p}_\perp},p_0,q_0,Q^2)
g({\bf {p}_\perp}) + g^2C_FT   \nonumber \\
&&\int \frac{d^2{\bf {\bf\ell}_\perp}}{(2\pi)^2}C({\bf {\ell}_\perp})\left[g({\bf {p}_\perp})-g({\bf{p}_\perp+\bf{\ell}_\perp})\right]~~~~~
\label{agmz-l}
\end{eqnarray}
In  above  equations, $r_0=p_0+q_0$ and ${\bf{f}},g$ are actually functions
of  $p_0,q_0,Q^2$ represented as, ${\bf {f}}({\bf {p}_\perp},p_0,q_0,Q^2)$,
$ g({\bf {p}_\perp},p_0,q_0,Q^2)$. 
The  Eq.\ref{agmz-t}  and  Eq.\ref{amy}  are
identical except for $\delta E$ energy factor. Further, Eq.\ref{agmz-t} and
the  Eq.\ref{agmz-l} are also similar except for the left
side of AGMZ equation  is a constant $\sqrt{|p_0r_0|}$.  
Aurenche {\it et. al.,} solved these equations, based on a very elegant method of impact parameter
representation \cite{lpmdilep}. As  mentioned
earlier, we have solved the AMY equation by the variational method and a new
method  called  iterations method. These methods are formulated in terms of
tilde variables. Therefore, it is  required  to  transform  the  two  above
equations to tilde quantities. Following the definition of \cite{arnold2} we have,
\begin{equation}
{\bf    \tilde{p}_\perp}=\frac{\bf p_\perp}{m_D}
~;~{\bf \tilde{f}(\tilde{p}_\perp)}=\frac{m_D}{T}{\bf{f}({p}_\perp)}~;~
{\tilde{\delta E}}({\bf\tilde{p}_\perp})=\frac{T}{m_D^2}{\delta E(\bf{p}_\perp}).
\end{equation}
One   can   see   that   the   collision   kernel   are   related   by  $
\tilde{C}{\bf(\tilde{\bf\ell}_\perp})=T  C({\bf\ell_\perp}).$
\begin{eqnarray}
2{\bf \tilde{p}_\perp}&=& i \delta \tilde{E}({\bf \tilde{p}_\perp},p_0,q_0,Q^2)
{\bf \tilde{f}}({\bf \tilde{p}_\perp}) \nonumber \\
&& + \int\frac{d^2{\bf \tilde{\bf\ell}_\perp}}{(2\pi)^2} \tilde{C}({\bf \tilde{\ell}_\perp})
\left[{\bf \tilde{f}}({\bf \tilde{p}_\perp}) -{\bf \tilde{f}}({\bf \tilde{p}_\perp+\tilde{\bf\ell}_\perp})\right]
\label{agmz-t-tilde}
\end{eqnarray}
\begin{equation}
\delta \tilde{E}({\bf \tilde{p}_\perp},p_0,q_0,Q^2)=
\frac{q_0T}{2p_0(q_0+p_0)}\left[\tilde{p}_\perp^2+\kappa
_{\mbox{\small{eff}}}\right]
\end{equation}
In  above
$\kappa_{\small{\mbox{eff}}}=\frac{M^2_{\mbox{\small{eff}}}}{m_D^2}$.    In
the equations below, we introduce the calligraphic symbol  $\cal{C}$  which
is  is a linear operator and this should not be confused with the collision
kernel $C(\bf{p_\perp})$.  The  integral  equations  can  be  put  in  the
following linear operator form.
\begin{eqnarray}
2{\bf \tilde{p}_\perp}&=&\left( i\delta \tilde{E}({\bf \tilde{p}_\perp})+ \tilde{\cal{C}}({\bf \tilde{p}_\perp})\right){\bf \tilde{f}}
\label{amyopeq}
\end{eqnarray}
\begin{eqnarray}
\tilde{\cal{C}}({\bf \tilde{p}_\perp}){\bf \tilde{f}}&=& \int\frac{d^2{\bf \tilde{\bf\ell}_\perp}}{(2\pi)^2}\tilde{C}({\bf \tilde{\ell}_\perp})
\left[{\bf \tilde{f}}({\bf \tilde{p}_\perp})-{\bf \tilde{f}}({\bf \tilde{p}_\perp+\tilde{\bf\ell}_\perp})\right]~~~~~
\label{amyop}
\end{eqnarray}
The  linear  operator  at  a  given momentum
${\bf{p_\perp}}$ gives the sum of  the 
transverse vector function  differences between the
current momentum and all other momenta arising from scattering   weighted by
collision kernel. The   above   linear   operator  suggests  that  the
corresponding tilde operator is related to untilded operator exactly as  the
$\delta  E$  operator.  It  can  be easily verified that for the transverse
equation            this            means,            $\tilde{\cal{C}}({\bf
\tilde{p}_\perp})=(T/m_D^2)\cal{C}({\bf     p}_\perp)$.
In the  AGMZ equation  when
$\delta E$ transforms to tilde quantity, its $T/m_D^2$  factor  has  to  be
absorbed  by  the transformation of $g$ function. Therefore, if we set the
$g$ to  transform  as  $\tilde  g({\bf  \tilde{p}_\perp})=  \frac{m_D^2}{T}
g({\bf  p_\perp})$, the equation for the longitudinal part could be written as,
\noindent
\begin{eqnarray}
2{\sqrt{|p_0r_0|}}&=& i \delta \tilde{E}({\bf \tilde{p}_\perp},p_0,q_0,Q^2)
{\tilde{g}}({\bf \tilde{p}_\perp}) + \nonumber \\
&&\int\frac{d^2{\bf \tilde{\ell}_\perp}}{(2\pi)^2}\tilde{C}({\bf \tilde{\ell}_\perp}) \left[{\tilde{g}}({\bf \tilde{p}_\perp}) -{\tilde{g}}({\bf \tilde{p}_\perp+\tilde{\bf\ell}_\perp})\right] ~~~~~
\label{agmz-l-tilde}
\end{eqnarray}
This implies that the  $\tilde{g}$
is  larger  than  $g$ by a factor $m_D^2$. Therefore, when the solution for
this Eq.\ref{agmz-l-tilde} is obtained and integrated over   $\int\frac{d^2\bf
\tilde{p}_\perp}{(2\pi)^2}\Re{\tilde{g}{\bf    (\tilde{p}_\perp)}}$,    the
result will be larger than the true result by exactly $m_D^2$ factor.  This
problem    was    not    present    for    the   transverse   part   because
$\bf{p_\perp.f(p_\perp)}$ is independent of any  $m_D^2$ factor arising from the tilde transformation.
Therefore,  to  take
this  anomaly  into  account correctly, we will introduce $\frac{1}{m_D^2}$
factor for the longitudinal contribution while estimating  the  imaginary  part  of
photon  polarization  tensor.  Further, in  the  tilde  variables, the above equation is
not   dimensionless.  Therefore,  we  will  divide  the  above
Eq.\ref{agmz-l-tilde}, by $m_D$ in order to get the following equation, where
 absorbing 1/$m_D$ factor, $g$ is now re-defined.
\begin{eqnarray}
2\sqrt{\frac{|p_0r_0|}{m_D^2}}&=& i\delta \tilde{E}(({\bf \tilde{p}_\perp},p_0,q_0,Q^2)
{\tilde{g}}({\bf \tilde{p}_\perp}) + \nonumber \\
&& \int\frac{d^2{\bf \tilde{\ell}_\perp}}{(2\pi)^2}\tilde{C}({\bf \tilde{\ell}_\perp}) \left[{\tilde{g}}({\bf \tilde{p}_\perp}) -{\tilde{g}}({\bf \tilde{p}_\perp+\tilde{\bf\ell}_\perp})\right] ~~~~~
\label{agmz-l-tilde-mod}
\end{eqnarray}
In the above equation,  $g$ transforms as
$\tilde g({\bf \tilde{p}_\perp})= \frac{m_D}{T} g({\bf p_\perp})$, similar to $\bf f(p_\perp)$ function. 
Therefore,  the anomalous factor is now  $\frac{1}{m_D}$.
In the present work, we have solved the above Eq.\ref{agmz-l-tilde-mod} by iterations and variational method.
 The  variational  method  for  longitudinal part has been re-derived  \cite{svssymp}. 
We could get  good agreement of distributions form iterations method  
and variational results   and these details  will be discussed elsewhere.
Following these two methods, we obtained the $\bf p_\perp$ distributions
for  the  bremsstrahlung and {\bf aws} cases for both transverse and longitudinal
components. We considered five  photon  energy  values, for each energy ten  photon  momenta
and ten quark momenta, a total of 500 distributions for
transverse and longitudinal components of bremsstrahlung. Similarly the distributions were obtained for {\bf aws} case. 
All these were compared in both   iterations  and  variational  method. 
\section{Empirical  Analysis of the Solutions of AMY, AGMZ integral equations}
\noindent
In all the calculations that follow, we have used  two flavors, three colors and $\alpha_s$=0.30 and
T=1.0GeV. In Eqs.\ref{x0}-\ref{x3} we define four dimensionless variables used in the following work. Here one may notice
that the variable $x_1$ is the inverse of $x$ variable
used in \cite{svsprc} for real photon production.
$I^{b,a}_{T,L}$ are defined in the Eqs.\ref{itdef},\ref{ildef} which are obtained by
integrating the ${\bf p_\perp.f(p_\perp)},g({\bf p_\perp})$ distributions  ( in Ref.\cite{svsprc} there was a 2 factor extra
 in the definition of $I$ as given by ${\bf 2p_\perp.f(p_\perp)}$ ).
$I^{b,a}_{T,L}$ are the quantities required for imaginary part of polarization tensor and the  rate equations.
\begin{eqnarray}
x_0&=&\frac{|(p_0+q_0)p_0|}{q_0T  }  \label{x0} \\
x_1& = & x_0\frac{M_\infty^2}{m_D^2}  \label{x1} \\
x_2 &=& x_0\frac{Q^2}{q_0T}  \label{x2} \\
x_3& =& \frac{q_0T}{Q^2} \label{x3}   \\
I^{b,a}_{T}&=&\int \frac{\bf d^2\tilde{p}_\perp}{(2\pi)^2}  {\bf \tilde{p}_\perp \cdot\Re\tilde{f}(\tilde{p}_\perp)}
\label{itdef} \\
I^{b,a}_{L}&=&\int \frac{\bf d^2\tilde{p}_\perp}{(2\pi)^2}  \Re\tilde{g}({\bf \tilde{p}_\perp)} \label{ildef}
\label{ildef}
\end{eqnarray}
\par
\noindent
In the above equations superscript $b,a$ represent bremsstrahlung or $\bf aws$ processes depending on the $p_0$ value used.
The subscripts $T,L$ represent contributions from transverse ($\bf f(p_\perp)$) or longitudinal $(g(\bf p_\perp))$parts.
Using the iterations data, we  first obtained the integrated
values of ${\bf p_\perp}$  distributions as given by  $I^{b,a}_{T,L}$ of Eqs.\ref{itdef},\ref{ildef}.
We define the emission functions $g^{b,a}_{T,L}$ in Eq.\ref{gbatl} which are functions of
variables $x^{b,a}_{T,L}$. These $x^{b,a}_{T,L}$  variables  are given in Eqs.\ref{xbt}-\ref{xal}.
The $g^{b,a}_{T,L}$  are obtained from corresponding $I^{b,a}_{T,L}$ values by multiplying with
$c^{b,a}_{T,L}$ coefficient factors which are given in Eqs.\ref{cbt}-\ref{cal}.
Therefore, using the  $I^{b,a}_{T,L}$  obtained from iterations and variational data, 
we constructed all the four $g^{b,a}_{T,L}$ in Eq.\ref{gbatl}.
 It should be noted that the quantities $x^{b,a}_{T,L}$ and  $c^{b,a}_{T,L}$ 
listed in Eqs.\ref{xbt}-\ref{cal} are not definitions. They are the results of a hard search 
for  dynamical variables hidden in the solutions of AMY and AGMZ equations.
%%%%%%--------------------------------------------------------------------
\begin{figure}
\hspace{-1.5cm}
\includegraphics[height=11.cm,width=10.0cm]{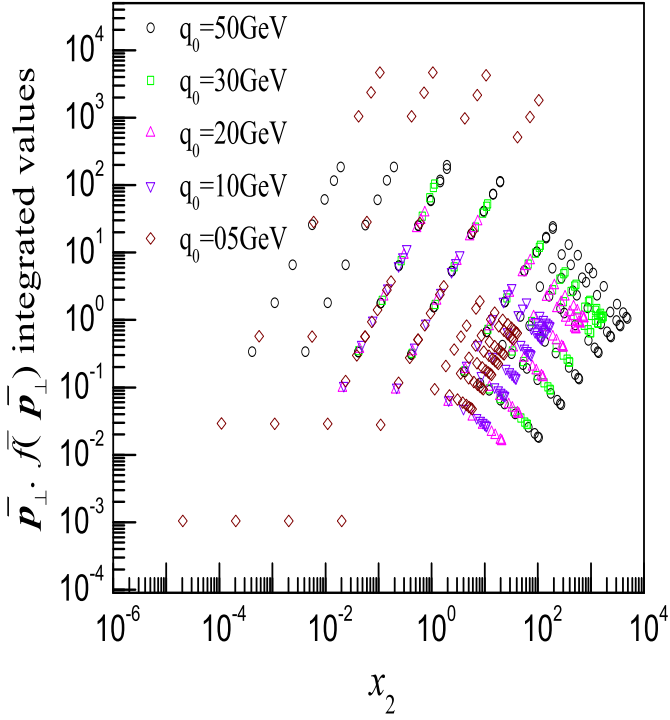}
\caption{  The  integrated values $I^b_T(x)$  versus dynamical
variable $x_2$ defined in the text.  The symbols represent the integrated values of
${\bf p_\perp}$ distributions of about 500 cases of $\{p_0,q_0,Q^2\}$ values.
The different colored symbols represent different photon energies.  }
\label{btp2int}
\end{figure}
%%%%%%--------------------------------------------------------------------
\begin{figure}
\hspace{-1.5cm}
\includegraphics[height=10.cm,width=10.0cm]{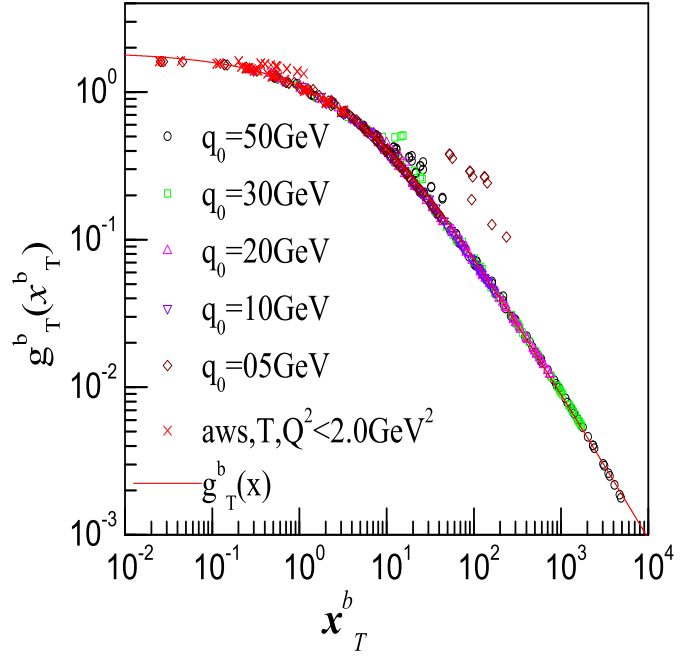}
\caption{  The  dimensionless  emission  function  $g^b_T(x)$  versus dynamical
variable $x^b_T$ defined in the text.  The symbols represent the integrated values of
${\bf p_\perp}$ distributions of about 500 cases of $\{p_0,q_0,Q^2\}$ values
and are transformed by suitable $c^b_T$  coefficient functions.
The different colored symbols represent different photon energies.  The red curve is an empirical fit given in the text.}
\label{btgx}
\end{figure}
%%%%%%--------------------------------------------------------------------
\begin{figure}
\hspace{-1.5cm} 
\includegraphics[height=10.cm,width=10.0cm]{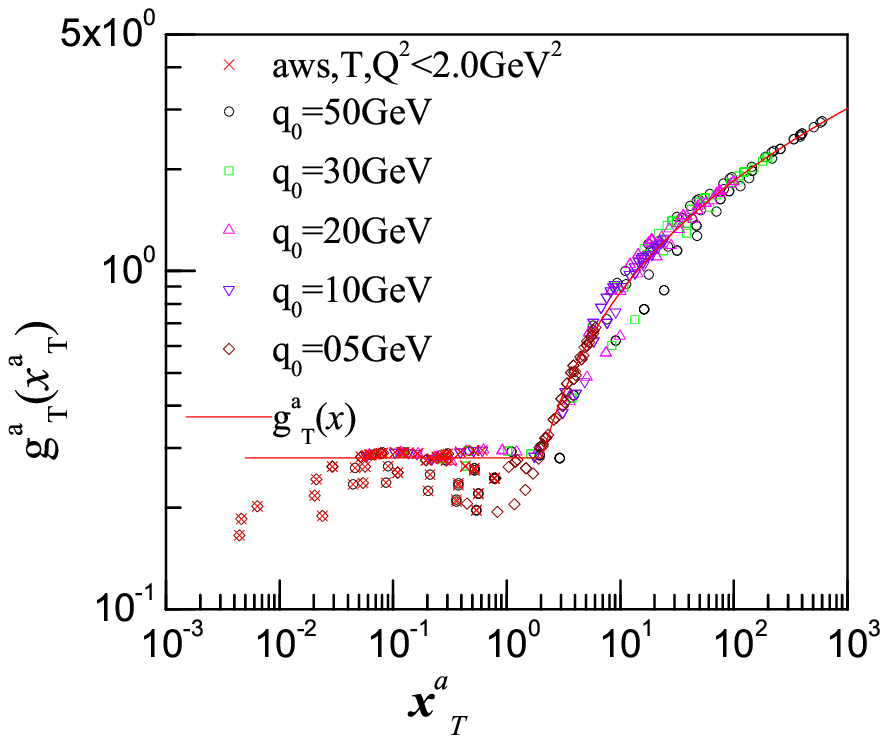}
\caption{  The  dimensionless  emission  function  $g^a_T(x)$  versus dynamical
variable $x^a_T$ defined in the text.  The required $c^a_T$  are the transformation coefficients
and  empirical fit (red curve) are given in the text.}
\label{atgx}
\end{figure}
%%%%%%--------------------------------------------------------------------
\begin{figure}
\hspace{-1.5cm}
\includegraphics[height=10.cm,width=10.0cm]{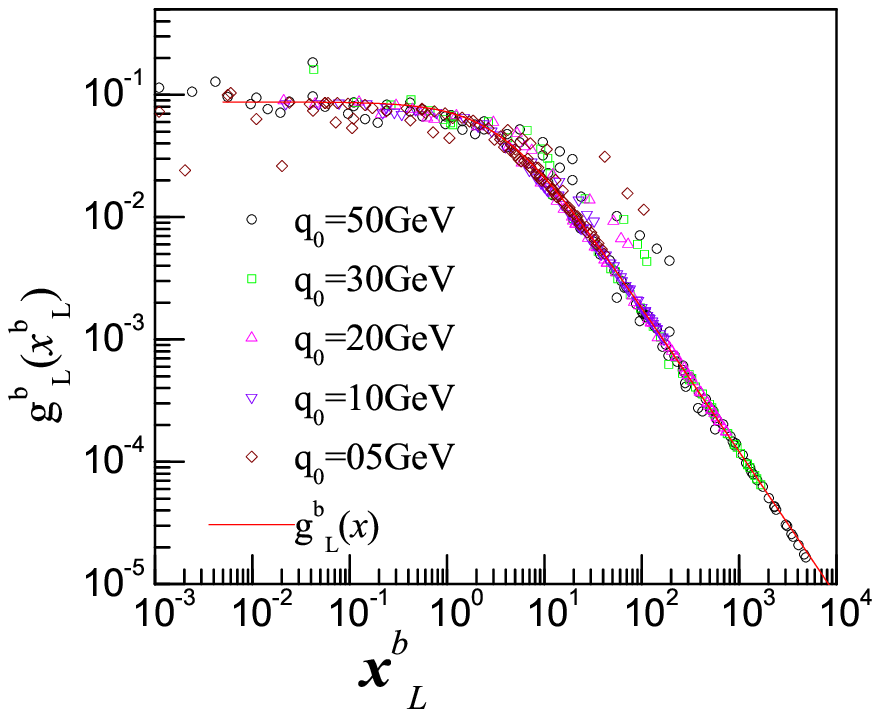}
\caption{  The  dimensionless  emission  function  $g^b_L(x)$  versus dynamical
variable $x^b_L$ defined in the text.  The required $c^b_L$  are the transformation coefficients
and  empirical fit (red curve) are given in the text.}
\label{blgx}
\end{figure}
%%%%%%--------------------------------------------------------------------
\begin{figure}
\hspace{-1.5cm}
\includegraphics[height=10.cm,width=10.0cm]{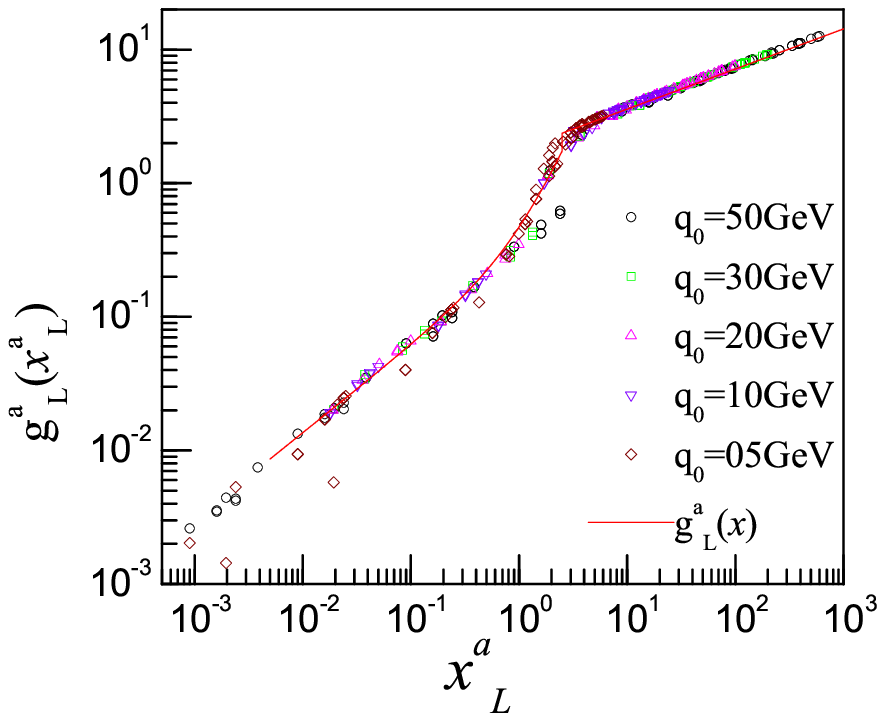}
\caption{  The  dimensionless  emission  function  $g^a_L(x)$  versus dynamical
variable $x^a_L$ defined in the text.  The required $c^a_L$  are the transformation coefficients
and  empirical fit (red curve) are given in the text.}
\label{algx}
\end{figure}
%%%%%%--------------------------------------------------------------------
\begin{eqnarray}
g^{b,a}_{T,L}(x^{b,a}_{T,L})&=&I^{b,a}_{T,L}(x^{b,a}_{T,L})c^{b,a}_{T,L}  \label{gbatl} \\
x^b_T&=&x_2+x_1   \label{xbt}  \\
x^a_T&=&x_2+0.18 x_1  \label{xat} \\
x^b_L&=&x_2      \label{xbl}    \\
x^a_L&=&x_2    \label{xal}
\end{eqnarray}
\begin{eqnarray}
c^b_T&=&\frac{1}{x_1^2}    \label{cbt}  \\
c^a_T&=&\frac{1}{x_1\left(1.0+x_3^{0.85}\right) x_2^{0.8}} \label{cat}  \\
c^b_L&=&\frac{Q^2}{(p_0(p_0+q_0))^{1.50}}\left(1.50+x_3^{0.75}\right)/x_2^{1/3} \label{cbl}     \\
c^a_L&=&\frac{1}{x_1^{1.4}q_0^{0.5}\left(1+x_3^{0.5}\right)}x_2^{0.2} \label{cal} 
\end{eqnarray}
Figure \ref{btp2int} shows the  $I^b_T(x)$ as a function of $x_2$ variable. All the bremsstrahlung transverse mode data consisting of
about 500 points are shown in the figure. The figure exhibits some trends as a function of  $p_0,q_0,Q^2$ values, but still does not bring out
the underlying dynamical scale hidden in the data. When plotted with respect to any other variables, the representation is much worse
and  no trend emerges from such plots. In  Figure \ref{btgx}, we show  the same data transformed to  $g^b_T(x^b_T)$ as a function of $x^b_T$ variable which is a
combination of $x_2$ and $x_1$ variables. The hidden scale is revealed in this plot, where data for all different $p_0,q_0,Q^2$  values  merged into a single curve.
Colored symbols represent  different photon energies. This is the transverse mode bremsstrahlung emission function.
Importantly, the crosses in the figure represent low $Q^2$ data for transverse mode of the ${\bf aws}$ process, intentionally 
plotted over here and will be  explained in the next figure. We have obtained an empirical fits to this data as given by $g^b_T(x)(empirical)$ formula in Eq.\ref{btgxemp}.
The need for the empirical fit will become clear later. 
\begin{eqnarray}
g^b_T(x)&=&\frac{10.0}{5.0+3.0\sqrt{x}+x}  \label{btgxemp} \\
g^a_T(x)&=&a+b x^p    \label{atgxemp} \\
a &=& -4.1395 ~;~  b = 4.1818 ~;~  p = 0.0779  \nonumber \\
g^a_T&=&0.28 ~~~\mbox{if}~~~ g^a_T(x)<0.28  \nonumber\\
g^b_L(x)&=&\frac{0.0876}{1+\left(\frac{x}{3.7727}\right)^{1.18}}  \label{blgxemp} \\
g^a_L(x)&=&0.2703 x^{0.65}+0.20 x^2    \label{algxemp} \\
\mbox{if}~ x&>&2.5 ~\mbox{then}~ g^a_L(x)=1.8 x^{0.3} \nonumber \\
I^{b,a}_{T,L}&=&\frac{g^{b,a}_{T,L}(x)}{c^{b,a}_{T,L}} \label{ibaemp}
\end{eqnarray}
%-----------------------------------------------------------------
\noindent
Similar exercise for transverse ${\bf aws}$ is shown in Figure \ref{atgx}. However, we notice that the $x$ variable is different from the
previous figure. Further, the $c^a_T$  that transform the integrated values into $g$ functions are very complex. We have parameterized this data
by an empirical function given in Eq.\ref{atgxemp}. One important thing to notice is  for low enough $Q^2<4.0GeV$ 
(possibly the data for $Q^2<4M_\infty^2$), the data deviates from general trends. These low $Q^2$
data is shown by crosses in the figure. It was noticed that the same data when transformed as required in Fig.\ref{btgx} and plotted versus  $x^b_T$,
they are very close to the bremsstrahlung function. This is shown by crosses in the previous  figure (Fig.\ref{btgx}).
Therefore, the transverse mode emission functions for the bremsstrahlung and for the low $Q^2$ ${\bf aws}$ are the same and are
given by $g^b_T(x^b_T)$(empirical). \\
The Figure \ref{blgx} shows the results for bremsstrahlung longitudinal component. The corresponding $c^b_L$ coefficient functions and the
empirical fits represented by $g^b_L(x)$ are given in Eqs.\ref{cbl},\ref{blgxemp}. Similarly, the Figure \ref{algx} shows the results for {\bf aws}  longitudinal component.
The corresponding $c^a_L$, $g^a_L(x)$ are given in Eqs.\ref{cal},\ref{algxemp}. \\
\begin{figure}
\hspace{-1.5cm}
\includegraphics[height=10.0cm,width=10.0cm]{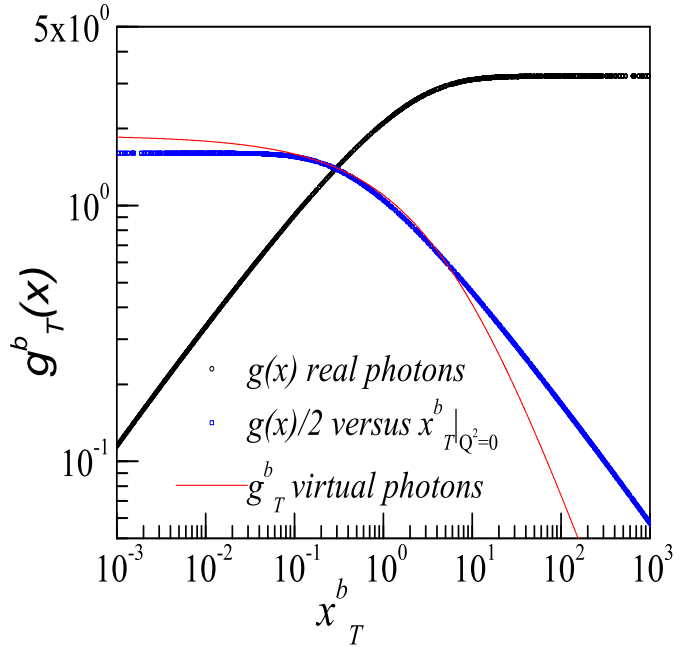}
\caption{  The  dimensionless  emission  function  $g(x)$  versus dynamical
variable $x$ taken from \cite{svsprc}, represented by black symbols. The half value of this
$g(x)$ in the inverse scale (that is same as present $x_1$ variable) is shown by blue symbols. 
This inverse variable coincides with the present
$x^b_T$ when $Q^2\rightarrow0$. The results for transverse mode of the virtual photon case
versus $x^b_T$ variable is shown by the red curve (same as in Fig.\ref{btgx}).}
\label{gxbtgx}
\end{figure}
\begin{figure}
\hspace{-1.5cm}
\includegraphics[height=10.cm,width=10.0cm]{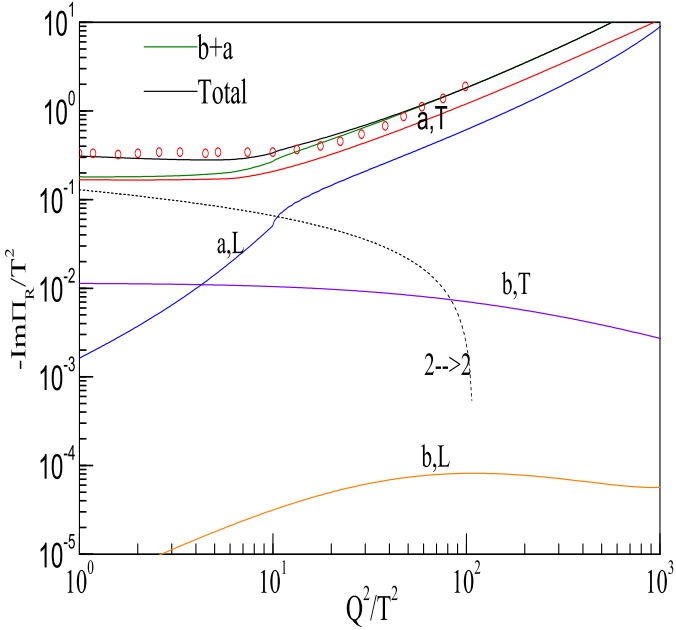}
\caption{$\Im \Pi_R$ plotted as a function of $Q^2/T^2$
 for a photon energy of 50GeV. The transverse components
of bremsstrahlung, $\bf aws$, the insignificant contribution from longitudinal parts,  and the
$2\rightarrow 2$ processes contribution using the Eq.\ref{oneloop} are all shown.
The purpose of the figure is to test  our empirical method compared to the results
of Fig.3 of Aurenche {\it et. al.,}\cite{lpmdilep} shown by red circles.}
\label{imself50md1}
\end{figure}
%%%%%%--------------------------------------------------------------------
\begin{figure}
\hspace{-1.5cm}
\includegraphics[height=10.cm,width=10.0cm]{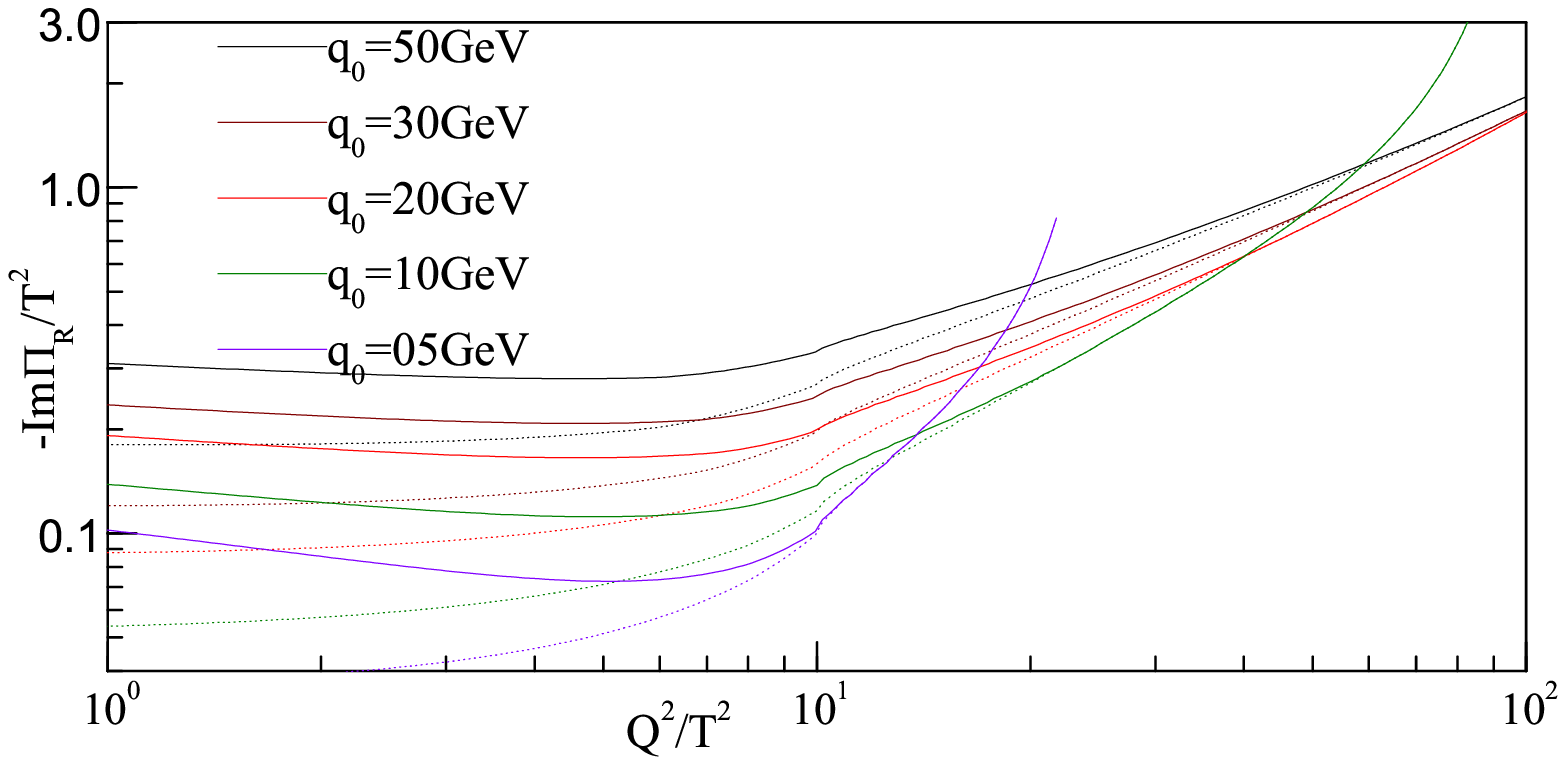}
\caption{Imaginary part of retarded photon polarization tensor for five photon energies represented by dashed curves 
in various colors obtained from the empirical emission functions. After adding the
$2\rightarrow 2$ contribution from Eq.\ref{oneloop}, the results are shown by colored solid curves.}
\label{imself2md1}
\end{figure}
%%%%%%--------------------------------------------------------------------
\noindent
In Figure \ref{gxbtgx}, we show the  photon  emission  function  in
terms of dynamical variable $x$ versus $g(x)$ defined and shown in Ref.\cite{svsprc}
for real photons.  We reproduce that data in the Figure. \ref{gxbtgx} represented by black symbols.
The inverse of this $ x$-variable of \cite{svsprc}
coincides with  the present $x_1$  value and is related to the  variables of the present study by
 $x^{-1}$(of Ref.\cite{svsprc})=$x^b_T\left|_{Q^2\rightarrow 0}\right.$.
This is the motivation for choosing the combination of $x_1$ and $x_2$ variables for transverse modes in the present study;~
as $Q^2\rightarrow 0$, the dynamical variables of the transverse parts go over to real photon dynamical variable.
Such a need is not there for longitudinal components.
Further, the $g(x)$   defined in \cite{svsprc} is two times larger than the  definition
in the present work for transverse modes (see factor 2 in  $2{\bf p_\perp.f(p_\perp)}$ in Eq.\ref{photrate} and not in Eqs.\ref{itdef},\ref{ildef}). 
Therefore the rescaled  $\frac{1}{2}g(x)$ versus $x^{-1}(=x^b_T\left|_{Q^2=0}\right.$)  for real photons is also plotted in the figure by blue symbols.
For comparison, the results of our present work for $g^b_T$ versus $x^b_T$  are also shown in the figure by red curve.
This red curve is the same red curve shown in Fig.\ref{btgx}, given by
$g(x)=10.0/\left(5+3\sqrt{x}+x\right)$.
It should be noted that in this figure, the variable for $x-axis$ is different for all the three curves.
Further, we explained earlier, that for   $Q^2<4.0GeV^2$, the transverse component of the $\bf aws$ data lies 
very close to the bremsstrahlung curve (see crosses in the figure \ref{btgx}).
From these results  it is obvious that for small quark momenta
and low virtuality of the photons emitted, the transverse part of the virtual photon ($\gamma^*$) emission function and real photon ($\gamma$) 
emission functions are the same as given by, $g^b_T(x^b_T) (\gamma^*) \approx  g(x) (\gamma)$ and this $x$ for real photons is simply
$x=x^b_T\left|_{Q^2=0}\right.$.
Further, the other longitudinal components for virtual photons are all negligible.
Therefore, for high photon energies and small photon virtualities, the virtual photon emission is identical to real photons emission
in the sense that their emission functions are the same. 
%%%%%%--------------------------------------------------------------------
\section{Empirical Estimation of Photon Retarded Polarization Tensor}
The imaginary part of retarded photon polarization tensor can be 
calculated using the $\bf p_\perp$ integrated values as shown in Eq.\ref{impolar}.
 In the previous section, we used the results from variational
 and iterations methods to obtain the $I^{b,a}_{T,L}$  values by integrating the distributions. 
We transformed these into $g^{b,a}_{T,L}$  functions shown in Figs.\ref{btgx}-\ref{algx}
and we fitted these by empirical functions  Eqs.\ref{btgxemp}-\ref{algxemp}.
The need for fitting arises,  for example, for any $\{p_0,q_0,Q^2\}$  values, 
we can generate the integrated values  $I^{b,a}_{T,L}(x)$
 as given by Eq.\ref{ibaemp}, without solving integral equations. 
The  imaginary part of retarded photon polarization tensor   (in units of $T^2$ and represented by $\Im \Pi_R$)  is calculated,  as  in  Eq.\ref{impolargx}.
Note that in this section,  no data from variational method or iterations is used.
The empirical fit functions provide the required results 
circumventing  the need to solve the transverse and longitudinal integral equations.     
The imaginary part of $\Pi_R$ has $\frac{Q^2}{q^2}$ factor for longitudinal part and this 
diverges as $Q^2\rightarrow q_0^2$. This trend is seen in the  Figs.\ref{imself50md1},\ref{imself2md1} as $Q^2$
increases\footnote{In this region, the formulae based on the kinematic conditions  used will not be valid, 
and that the divergence serves as a good signal reminding us of the non-validity of the 
formulae (as suggested  to me by F. Gelis).}. In this Eq.\ref{impolargx}, one should note the multiplying factor 
$\left(80\alpha_s\kappa/3\pi\right)$, in the present work it is equal to $\frac{2}{\pi}$. 
Importantly, one should note   the factor $\frac{1}{m_D}$   in the longitudinal part.
This is because the $\tilde{g}$ is already rescaled in Eq.\ref{agmz-l-tilde-mod}.  
In Eq.\ref{impolargx} the superscript $i$ denotes $ \{b,a\}$
bremsstrahlung and annihilation depending on the value of the integration variable $p_0$.
All terms in this equation are calculated and 
plotted  in Fig.\ref{imself50md1} to show the contribution to polarization tensor. 
\begin{eqnarray}
\Im{\Pi^\mu}_{R\mu} &=& \frac{e^2N_c}{2\pi} \int_{-\infty}^\infty  dp_0 [n_F(r_0)-n_F(p_0)] \otimes \nonumber \\
&& \int \frac{d^2{\bf \tilde{p}_\perp}}{(2\pi)^2}\left[ \frac{p_0^2+r_0^2}{2(p_0r_0)^2} \Re{\bf \tilde{p}_\perp.\tilde{f}(\tilde{p}_\perp)} + \right. \nonumber \\
&& \left. \frac{1}{\sqrt{\left|p_0r_0\right|}}\frac{Q^2}{q^2} \Re \tilde{g}({\bf\tilde{ p}_\perp})\right] 
\label{impolar}
\end{eqnarray}
\begin{eqnarray}
\Im{\Pi^\mu}_{R\mu} &=& \frac{e^2N_c}{2\pi} \int_{-\infty}^\infty  dp_0 [n_F(r_0)-n_F(p_0)] \otimes \nonumber \\
&&\left(\frac{2}{\pi}\right) \left[ \frac{p_0^2+r_0^2}{2(p_0r_0)^2}\left(\frac{g^i_T \left(x^i_T\right)}{c^i_T}\right)+ \right. \nonumber \\
&& \left. \frac{1}{\sqrt{\left|p_0r_0\right|}}\frac{Q^2}{q^2} \left(\frac{1}{m_D}\right) \left(\frac{g^i_L \left(x^i_L\right)}{c^i_L}\right)  \right] 
\label{impolargx}
\end{eqnarray}
The contribution of oneloop processes to imaginary part of the photon polarization  tensor  is
given by  equation below (formula taken from \cite{lpmdilep}). We will use this formula to evaluate the contribution of one loop processes 
and add to our empirical result to obtain the total result for retarded  photon polarization tensor. 
\begin{eqnarray}
Im{\Pi_R}^\mu_\mu\left|_{2\rightarrow2}\right. &=&-\frac{e^2g^2N_cC_FT^2}{16\pi} \left[ln\left(\frac{2q_0T}{Q^2}\right)  \right. \nonumber \\
&& \left. +1+\frac{ln(2)}{3}-\gamma_E+\frac{\zeta^\prime(2)}{\zeta(2)}\right]
\label{oneloop}
\end{eqnarray}
Figure \ref{imself50md1} shows the  $\Im \Pi_R$ plotted as a function of $Q^2/T^2$
 for a photon energy of 50GeV.  
This figure has been compared with the results of 
 Fig.3 of \cite{lpmdilep} represented by red symbols (approximately) for testing our empirical approach. 
It can be seen that the empirical approach has predicted rather remarkably well the the results of 
\cite{lpmdilep} over the $Q^2/T^2$~in the range of ~$1-100$. Further, the polarization tensor increases rapidly with $Q^2/T^2$ beyond 100.0 
The bremsstrahlung is insignificant obviously because the $q_0$ is very high. The transverse component of
$\bf aws$ only contributes, with a small contribution from longitudinal component above $Q^2>20GeV^2$.
The $2\rightarrow 2$ processes contribution is calculated using the Eq.\ref{oneloop}. 
The modification of the imaginary part of photon polarization tensor due to multiple re-scatterings  in the medium
has been well discussed in  \cite{lpmdilep}. The LPM effects only marginally increase the
$\Im \Pi_R$ at low $Q^2$. Here $2\rightarrow2$ processes also contribute. Importantly, the rescattering corrections  smooth out the discontinuity
at the threshold $Q^2=4M_\infty^2$. In order to understand how $\Im \Pi_R$  behaves with photon energy, we have calculated
this quantity using empirical approach for  five photon energies. These results are shown in Figure \ref{imself2md1}. The dashed curves
are the results from present calculations. When $2\rightarrow2$ contributions are added to this, we get the solid curves.
\section{Conclusion}
\noindent
The  photon  emission  rates  from  the  quark  gluon plasma have been
studied as a function of photon mass, considering LPM suppression effects
at  a fixed temperature. Self-consistent  iterations method and the variational method
have been used to solve  the  AMY and AGMZ  integral  equations.
We obtained  the ${\bf \Re\tilde{f}}({\bf    \tilde{p}_\perp})$, ${ \Re\tilde{g}}({\bf    \tilde{p}_\perp})$
distributions as a function of photon mass, photon energy and quark momentum.
The integrated values of these distributions for bremsstrahlung and $\bf aws$
show a very good scaling in terms of different dynamical variables for these two processes.
We defined four new dynamical variables $x^b_T,x^b_L,x^a_T,x^a_L$   for transverse and longitudinal components of bremsstrahlung
and  $\bf aws$ mechanism. In addition, we defined four new emission functions
namely $g^b_T(x^b_T)$,$g^a_T(x^a_T)$,$g^b_L(x^b_L)$$g^a_L(x^a_L)$,
that describe the photon emission for these four components.
We have obtained empirical fits to these
emission functions that have been constructed using the exact numerical calculations.
In terms of the new emission functions, estimation of the imaginary part photon polarization tensor
reduces to one dimensional integrals.
Using this empirical approach, we have calculated the imaginary part of retarded photon polarization tensor
as a function of photon energy and mass. These self energy graphs inlcude the LPM effects
due to rescattering in the plasma medium. For small photon $Q^2$ values and high photon energies, the
virtual photon emission function is a single function coinciding with the real photon emission function.
%%%%%%--------------------------------------------------------------------
\acknowledgements{I acknowledge fruitful discussions with Dr. Francois Gelis during the course of this work.
I thank Profs. Raghava Varma, Ajit K. Sinha, Drs. A.K. Mohanty, Alok Saxena, S. Ganesan and S. Kailas
for discussions. Computer Division is thanked for the computational services.}
\noindent
%%%%%%--------------------------------------------------------------------
%\begin{thebibliography}{99}

%\end{thebibliography}
%%%%%%--------------------------------------------------------------------
\end{document}